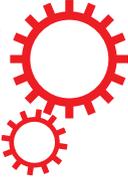

# Lithosphere strain rate and stress field orientations near the Alpine arc in Switzerland

N. Houlié[1,2], J. Woessner[3,4], D. Giardini[1] & M. Rothacher[2]

In this study we test whether principal components of the strain rate and stress tensors align within Switzerland. We find that 1) Helvetic Nappes line (HNL) is the relevant tectonic boundary to define different domains of crustal stress/surface strain rates orientations and 2) orientations of *T*- axes (of moment tensor solutions) and long-term asthenosphere cumulative finite strain (from SKS shear wave splitting) are consistent at the scale of the Alpine arc in Switzerland. At a more local scale, we find that seismic activity and surface deformation are in agreement but in three regions (Basel, Swiss Jura and Ticino); possibly because of the low levels of deformation and/or seismicity. In the Basel area, deep seismicity exists while surface deformation is absent. In the Ticino and the Swiss Jura, where seismic activity is close to absent, surface deformation is detected at a level of ~2 $10^{-8}$/yr (~6.3 $10^{-16}$/s).

**Motivation.** GPS networks, that measure long-term deformation at the surface, provide a unique opportunity to place seismic activity in context with geological observations resulting of long periods of tectonic activity (i.e. sea-level changes, geology, and erosion rates). Although they highlight various periods of times, the relationship between surface deformation (i.e. GPS), crustal seismicity (i.e. moment tensors principal components), and strain in the asthenosphere (i.e. shear wave splittings) is of genuine interest, both for understanding the origin of the surface deformation observed and quantifying seismic hazard. For instance, the comparison of directions of principal components of both stress and strain tensors helps understanding whether stress accumulates along faults[1–5] or through blocks[6–8], whether seismic ruptures are triggered[9–15], encouraged or impeded[16] and, finally, how stress is transferred to surrounding fault systems[17–20]. Also, the comparison between strain rates and asthenosphere strains sheds light on the upper mantle strength[21,22] and reveals whether the current tectonic period could be responsible and/or parented to the anisotropy detected in the mantle[23–25].

Across strongly active plate boundaries, linking seismic activity and surface deformation is made easier because strain rates are high ($>10^{-7}$/yr) and permanent GPS networks are dense[26–28]. There, higher strain rates tend to shorten seismic cycles lengths[27], enabling us to document larger portions of the seismic cycle and informing us on the preparation of future mainshocks. For slowly deforming intra-plate regions, however, we understand less about the relationship between surface deformation and seismicity: strain rates are smaller, seismic activity may not be sufficient to deform the surface while seismic hazard may still be significant[29–32]. We take Switzerland for an example of this situation and investigate the link between GPS strain rates and the seismicity observed since 2 decades.

**Seismic Activity.** At the meeting point between the central European platform and the Adria plate (Fig. 1a), Switzerland is at a key location to understand the dynamics of lithosphere in the region. Over the last millennium, historical earthquakes such as for $M_w = 6.6$ Basel 1356[33–36], $M_w = 6.2$ Churwalden 1295[37,38], $M_w = 6.1$ Visp 1855 Valais[39–44] and $M_w = 6.1$ Sierre 1946[45,46] were reported. Since 1500's six $M_w \sim 6.0$ events in the wider Valais area[47] ruptured different fault systems with an average recurrence interval of ~100 years[48]. Another peculiarity of Switzerland's seismicity lies in its depth distribution (Fig. 2a). North of the Helvetic Nappes (*HN*), hypocentre depths are limited to the upper crust with extreme values up to Moho depths (~35 km) near Luzern, while south of the *HN*, the earthquakes depths remain shallower than 15 km[49,50]. The transition from shallow to deep seismicity was interpreted as an indication of high pressure fluids[51]. Such a transition (topography, geology) between the Swiss Molasses and the Alpine arc domain cannot be confirmed using the moment tensor catalogue: most focal mechanisms (first motions solutions) are of strike-slip type except in the southern Valais (Fig. 2b). This last observation raises questions regarding the tectonic status (active, non-active) and real nature of the Alpine arc's current dynamics (e.g. resulting of gravitational spreading, surface expression of a lithosphere discontinuity).

[1]ETH-Zurich, SEG, Sonneggstrasse 5, 8092, Zürich, Switzerland. [2]ETH-Zurich, MPG, HPV G 53, Robert-Gnehm-Weg 15, 8093, Zürich, Switzerland. [3]ETH-Zurich, SED, Sonneggstrasse 5, 8092, Zürich, Switzerland. [4]Present address: Risk Management Solutions (RMS), Stampfenbachstrasse 85, CH-8006, Zurich, Switzerland. Correspondence and requests for materials should be addressed to N.H. (email: nhoulie@alumni.ethz.ch)

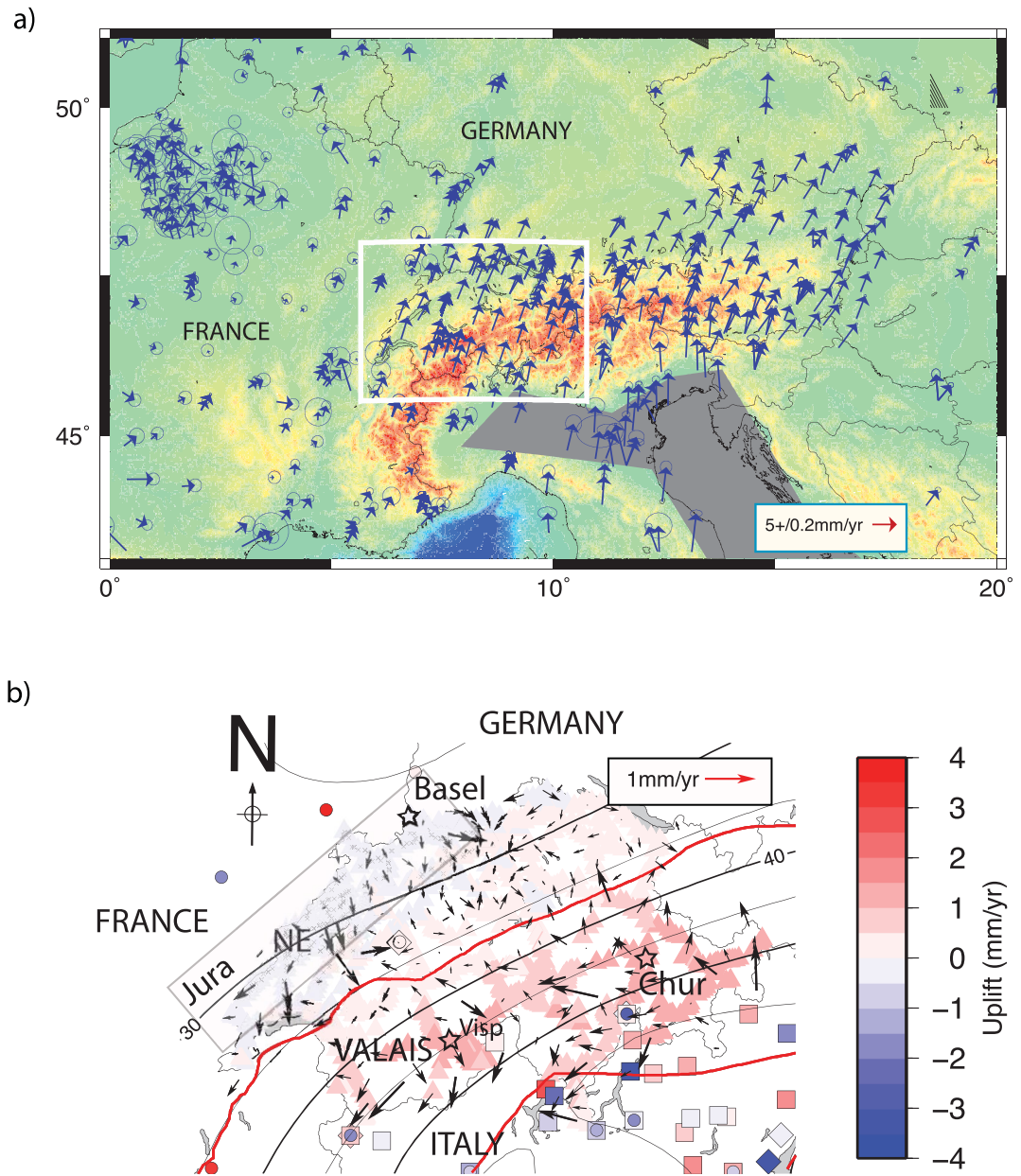

**Figure 1.** (**a**) Regional GPS velocity field (blue, stable Eurasia fixed) in western and central Europe in the vicinity of the Alpine arc and of the Adria plate. (**b**) Comparison of relative GPS horizontal velocities (black arrows), SWISSTOPO levelling (triangles) vertical rates[70] and GPS vertical rates[74,96] (squares and diamonds correspond to Serpelloni *et al.*, 2013 and Cenni *et al.*, 2013's measurements, respectively). Levelling rates and GPS vertical rates in Italy are indicated with color-coded triangles and color-coded squares. All geodetic motions are plotted with respect to the site ZIMM (circle inside a square). Moho depths (km)[97,98] are overlaid using black lines (one line per 5 km). A reasonable correlation between Moho depth and highest uplift rates (>1 mm/yr) is visible. NE stands for Neuchatel. HNL and Insubric lines are indicated in red and green, respectively. This figure has been created using Generic Mapping Tools 4/5[99] and Adobe Illustrator CS3.

**Surface Deformation.** The surface deformation of the region is limited[52–54], and is considered to be in agreement with moderate-to-low level of seismic activity. Horizontal strain rates ($<10^{-7}$/yr) are likely the consequence of the motion of the Adria plate[55,56] towards the North-East[57,58]. Because of current shortening rates are small (<3 mm/yr[59]) and knowing the total convergence is from 300[60] to 480 km[61], the Alpine belt is today seen either as inactive[62,63] and/or close to isostatic equilibrium[64]. The latter is in agreement with vertical rates (up to 1.2 mm/yr south of the Helvetic Nappes) observed by both multi-decadal levelling[65,66] and GPS measurement campaigns[67–70]. Because uplift rates are densely mapped in Switzerland, such a whole-arc view on vertical motions can however be refined. For instance, there is no correlation between elevation of the topography and uplift rates inferred of levelling campaigns, except in the Jura arc (Fig. 3). There, the uplift rates are in agreement with levelling measurements made in the Rhine graben[71] or in the Jura very locally[72], but are also anti-correlated with elevations in Jura at the arc scale (Fig. 3). On the horizontal plan, the situation is also complex, reflecting the interplay between inherited geology[73] and active processes. Together, horizontal convergence and uplift variations[74] explain why, even if they remain rare, large magnitude ($M_w > 6.0$) earthquakes occur[37].

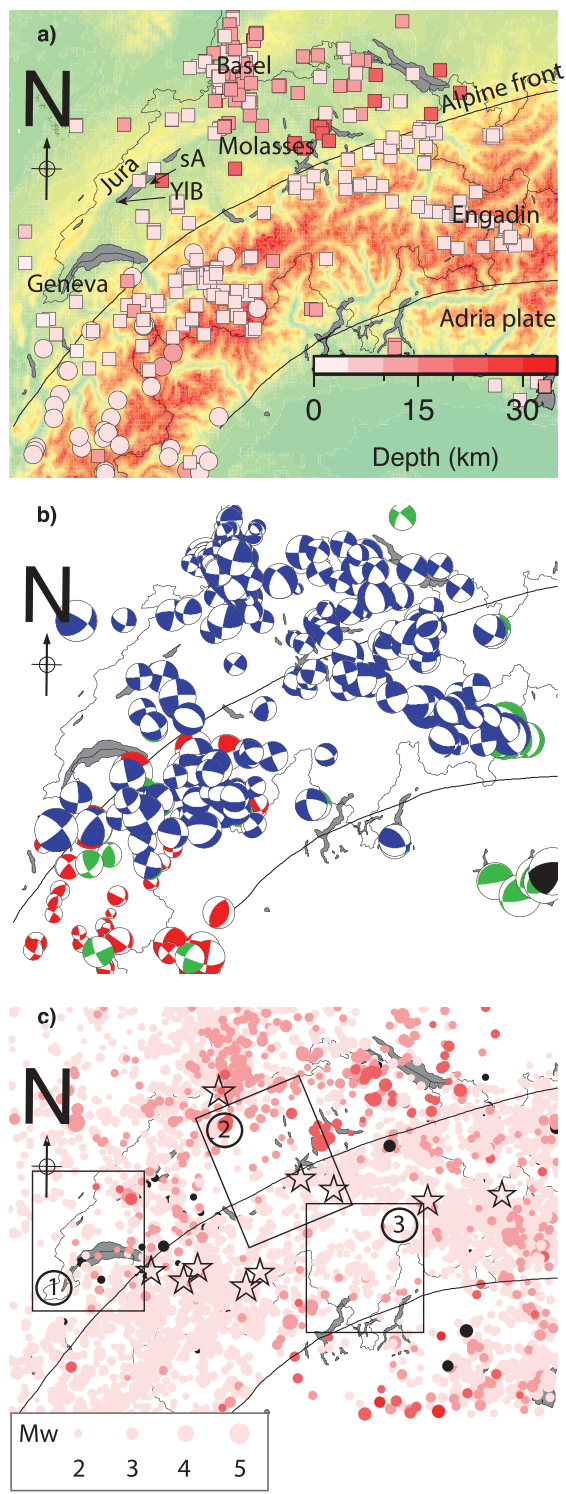

**Figure 2.** Seismicity characteristics. (**a**) Depth of earthquakes hypocenters. If we exclude the Geneva area, largest earthquakes are deeper north of the HNL (>12 km). The transition between deeper and shallower seismicity is visible across the HNL where the uplift rates are the highest. sA and YlB stand for St-Aubin and Yverdon-les-bains. (**b**) First-motion focal mechanisms[49,81,94,100,101]. [Green: International Seismological Center (ISC); Red: Sue and Delacou *et al.*, 2004; Blue: Kastrup *et al.*, 2004 + Deichmann *et al.*, 2012 + Marschall *et al.*, 2013; Black: QRCMT INGV]. Most events mechanisms are close strike-slip type suggesting that the orientations of principal stress axes are located in the horizontal plane ($\sigma_3$ nearly vertical). Helvetic Nappes (*HN*) and Adria plate boundaries are drawn (black lines)[102]. (**c**) ECOS-09 seismicity catalogue[37]. This figure has been created using Generic Mapping Tools 4/5[99] and Adobe Illustrator CS3.

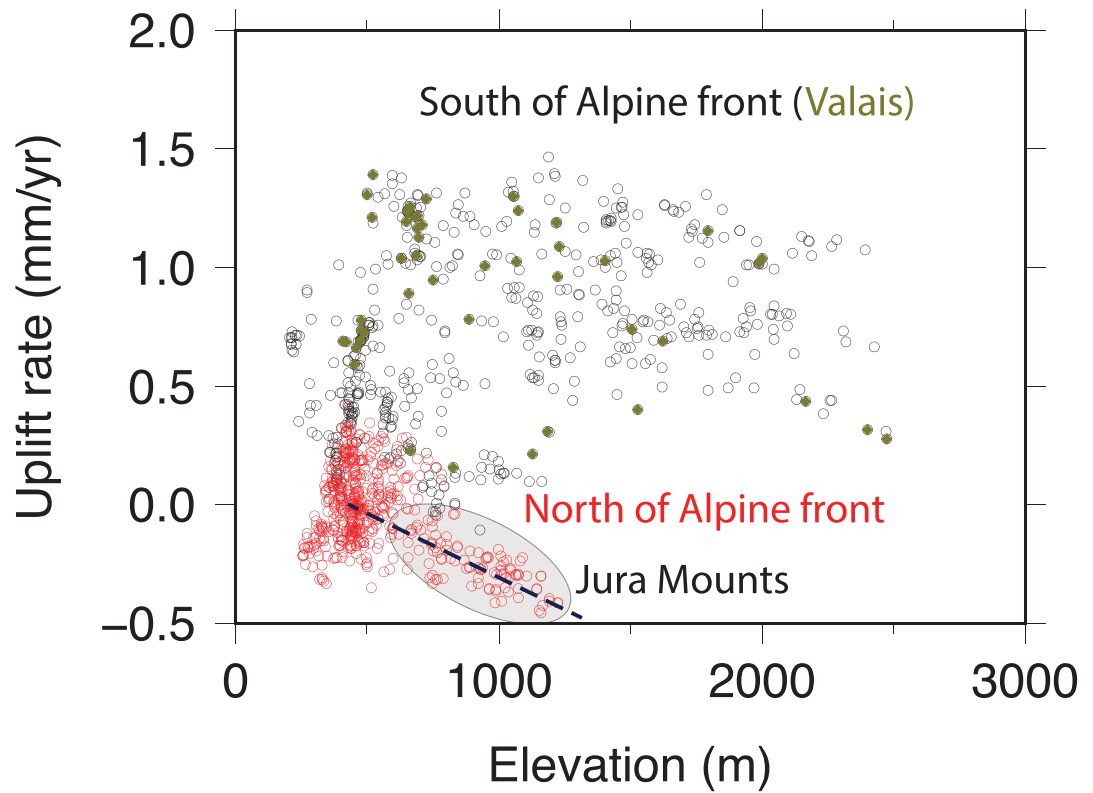

**Figure 3.** Uplift rates obtained from repeated levelling campaigns as a function of the elevation. No correlation is found between rates and elevations except in the Jura (grey ellipse) where highest points are also the fastest subsiding ones. Levelling vertical rates[70]. and topography from SWISSTOPO (200 m resolution). This figure has been created using Generic Mapping Tools 4/5[99] and Adobe Illustrator CS3.

In order to understand the tectonics acting in the central Alps in Switzerland, we 1) map the strain rate field measured using GPS during the last 2 decades, 2) investigate whether the *HN* can be associated with changes of orientations of the principal components of the strain tensors and 3) at last, by comparing those with orientations of *P-/T-* axes, and fast axes of shear waves splittings, we discuss whether the orientations of principal components of the stress tensors constrained by moderate magnitude seismic events ($M_l > 2.0$) are compatible with mantle deformation observed in the region[63].

### Results

At the country scale, shear and extension rates dominate compression rates (Fig. 4a–d). This observation fits well with a crustal seismicity mostly composed of strike-slip events and with previous studies focusing on long-term deformation of central and western Alps[75,76]. Spatially, the *HN* cannot be seen as a limit across which strain rates vary. Indeed, in the region of Lausanne and Geneva (shear rates up to ~4.0 $10^{-8}$ strain/yr, Fig. 4c) and within in the Jura arc (~2.0 $10^{-8}$ strain/yr), deformation is detected. Unlike seismicity (Fig. 4c), strain rates are spatially homogeneous. However, such a crude view of Switzerland's surface deformation is quickly challenged by comparisons of seismicity and strain rates performed at a more local scale.

In specific areas, we find apparent discrepancies between strain rates and seismic activity (i.e. occurrence of earthquake).

- In the central Jura, ~NW-SE extension is found (Fig. 4a) while only low seismicity is observed. This observation appears to be surprising although it is not the first time that extension was constrained across the Jura arc: the same conclusion was reached after mapping the strain field of Switzerland using 3D interpolation of strain rates on a regular grid[77]; and after a joint processing of the AGNES/COGEAR datasets[78]. The extension is therefore not due to an artefact of network geometry or an interpolation issue. Due to the poor density of the network in the region, extension across the Neufchatel lake is not the unique mechanism that is able to explain the GPS vector field in this area. The same vector field could be generated by the motion associated with a strike slip fault oriented NW-SE and located between the localities of St Aubin and Yverdon-les-bains. Such faults have been documented in the area[79], but further investigation will be necessary to identify which tectonic structure may be responsible for the deformation observed.

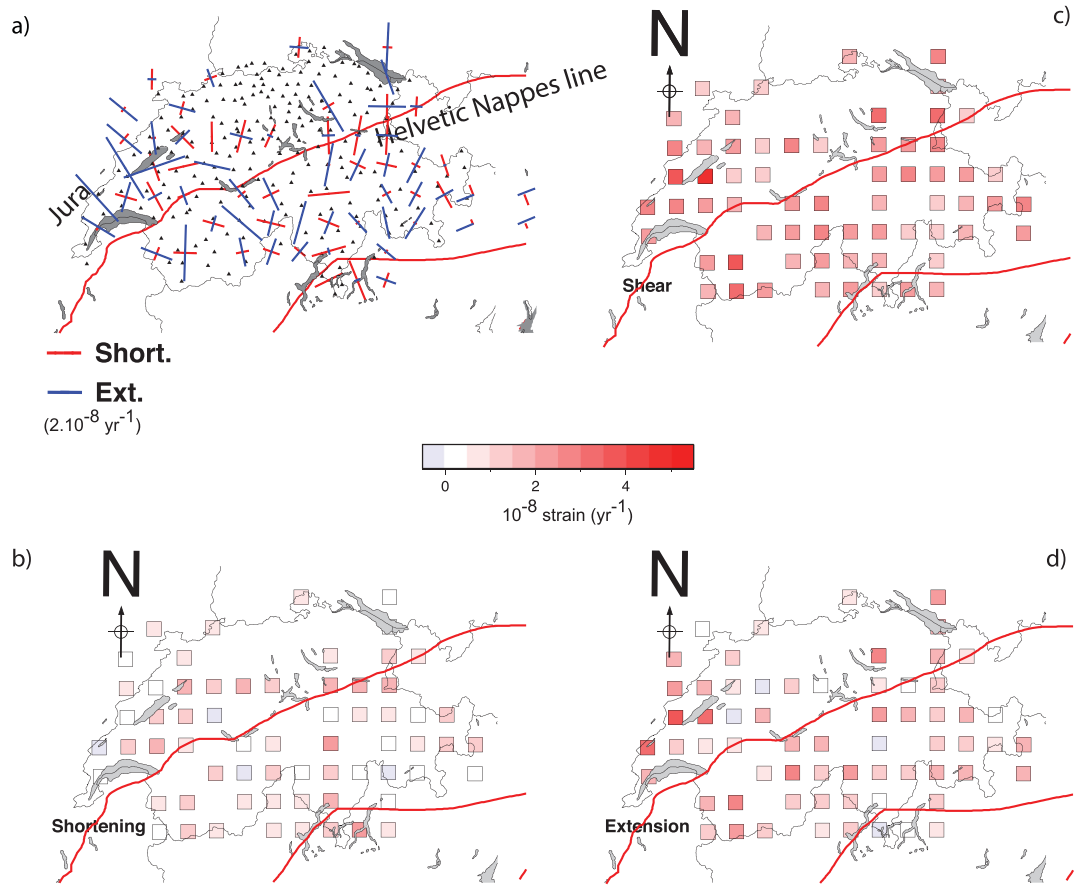

**Figure 4.** Horizontal strain field of the central Alps and Switzerland computed the inversion of GPS data. Alpine front and Adria plate are indicated using red lines. (**a**) Principal strain rate axis (compression in red, extension in blue) computed using the SSPX 2D algorithm[95]. Strain rates have been estimated using a near-neighbour smoothing strategy (grid spacing 25 km, 6 neighbours, maximal distance = 50 km). Shear and extension dominate the strain field in the whole country. Some areas such as Luzern, Zurich and Basel remain un-deformed even if they are the places of active seismicity; (**b**) magnitude of maximum shortening rates (**c**) maximum shear strain rate and (**d**) Maximum extension rate. Basel (BA), Geneva, (GE), Luzern (LU), Konstanz (KS), Zurich (ZU), Valais (VA) and Ticino (TI) are indicated on panel b. Average shear, extension and shortening strain rates are respectively equal to 2.2 ($\pm$0.8) $10^{-8}$/yr, 1.4 ($\pm$0.10) $10^{-8}$/yr and $-7.7$ ($\pm$0.6) $10^{-8}$ /yr. This figure has been created using Generic Mapping Tools 4/5[99] and Adobe Illustrator CS3.

- South of the *HN*, directions of maximum shortening are parallel to the front, (i.e. ~E-W direction) except in the east near Engadin where shortening directions are ~N-S (Fig. 8b). This observation is well in agreement with observation made across the Italy-Switzerland border[80].
- In the north-western Valais, the ~N-S extension (few ~E-W compression) is in agreement with the focal mechanisms (Fig. 2b) inferred from first motion arrivals[81].
- In the north-east of the Swiss Molasses, no substantial deformation is observed where diffuse seismicity is still observed (Fig. 2c). In the Ticino, both uplift (~1.0 mm/yr) and deformation (2.0 $10^{-8}$/yr) are observed while $M_W > 4$ earthquakes are absent in historical seismicity catalogues of this area[39]. There, the GPS network is not dense enough to capture local deformation.

In a situation in which strain rate is detected but earthquake magnitudes are too small to complete a moment tensor analysis, we attempt to use the strain rate field to characterize the regional stress field through the orientation of principal stress component. Indeed, it has been shown[21] that the directions of strain rate tensor components inferred from GPS are a good proxy of the principal stress components projected in the horizontal plane (i.e. ~orientation of $S_H$). Following the same idea, orientations of principal components of the strain rate tensor components with borehole breakouts orientations were successfully compared[22]. Here, we compare directions of the maximum shortening rates (GPS) with *P*-/*T*- axes orientations in order to constrain the orientation of $\sigma_1$. Comparison of orientations of $S_H$ and *P*- axes have been used in the past to infer the direction of $\sigma_1$ at the local scale[82–84] within uncertainties[85]. The alignment of maximum shortening directions with $S_H$ and *P*- orientations would imply that the lithosphere strain rates and crustal stress fields are consistent. First, we find that strain and stress components orientations are similarly scattered (~20 degrees; Table 1) both north and south of the HN. Interestingly, shortening strains, *T*- and *P*- axes orientations rotate across the HN of +32, −28 and −15 degrees N, respectively (Table 1). Rotation is well visible for the strain principal axes along defined profiles but also within groups (Figs 5, 6 and 7). Everywhere, the orientations of maximum shortening and of *P*- axes fall in the same quadrant (Fig. 8) and are associated with the location of HN (or to the Basal Alpine thrust). Such an agreement suggests that 1) south of the *HN* the maximum shortening axes directions can be considered a good proxy for the direction of the principal stress component and therefore 2) lithosphere deformation and crustal seismic activity are consistent. The spatial consistency of the rotation for strain/stress datasets, nevertheless of their inherent diversity and of their lack of causality (it is difficult to understand how one induces the other), implies that both strain-rate and stress fields are induced by a same set of forces deforming the crust and possibly the lithosphere at the regional scale. From this spatial analysis we conclude that the HN delimit spatially two regimes of deformation and seismic activity. Variation of the strain rates across the Alpine arc is not visible, suggesting the strain is well distributed across fault systems, as observed in the seismicity catalogue (Fig. 2c).

|   | All CH | North of Alpine front | South of Alpine front | Reference |
|---|---|---|---|---|
| Shortening | 81 +/− 49 | 65 +/− 49 (N = 24) | 97 +/− 38 (N = 37)* | This study |
| Extension | −8 +− 49 | 7 +/− 1 (N = 12) | −5 +/− 49 (N = 21) | This study |
| SKS | 39 +/− 17 | 39 +/− 17 (N = 8) | 38 +/− 17 (N = 14) | 63 |
| P-Axes | 151 +/− 38 | 158 +/− 32 (N = 110) | 143 +/− 43 (N = 95) | 94 |
| T-axes | 41 +/− 32 | 54 +/− 31 (N = 110) | 26 +/− 25 (N = 95) | 94 |
| P-axes** | 153 +/− 28 | N/A | N/A | 104 |
| T- axes** | 66 +/− 18 | N/A | N/A | 104 |

**Table 1.** Azimuths (deg. N) of T- and P- axes, directions of maximum shortening, extension and SH. *We excluded the 5 points located south of the Alpine front but deformed by principal strain shortening in the NS direction (circle on Fig. 8c). **Only events located within 12 km above the Moho are considered here.

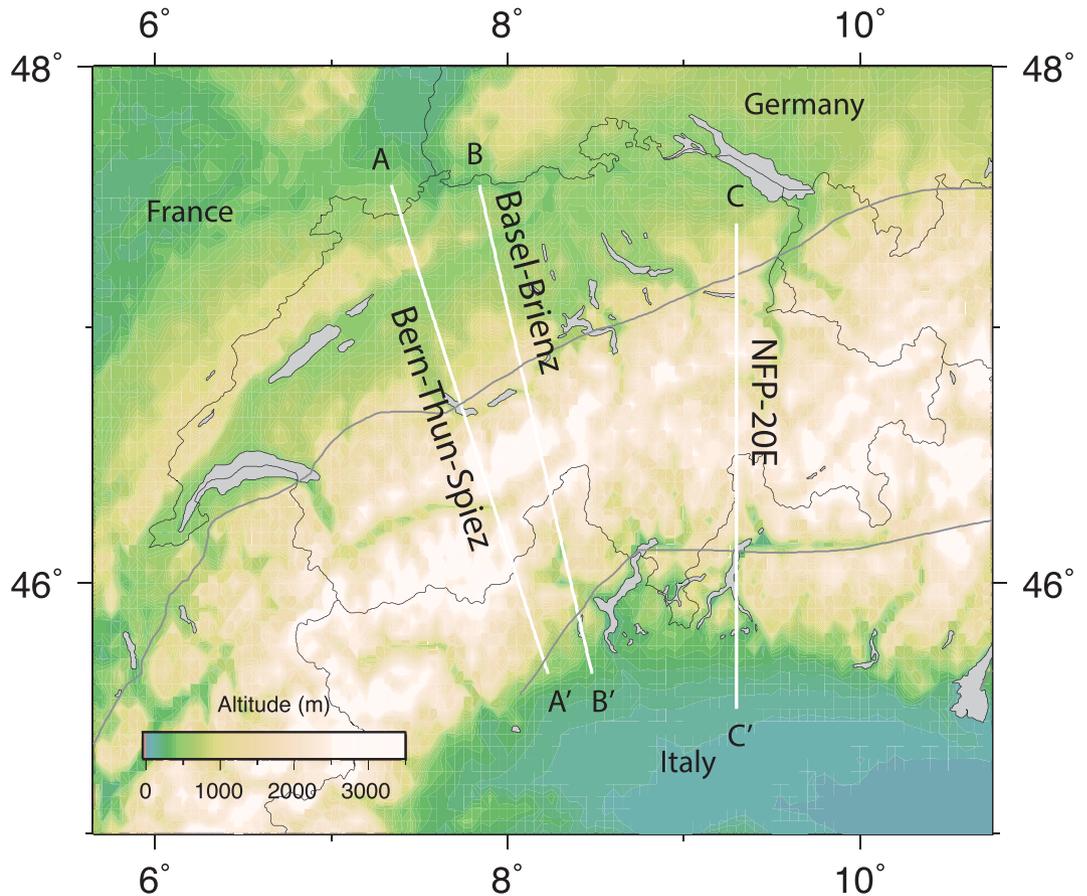

**Figure 5.** Map showing the locations of profile used in Figs 6 and 7. This figure has been created using Generic Mapping Tools 4/5[99] and Adobe Illustrator CS3.

Regarding the balance between long- and short-term deformation, we complete a comparison between shear wave splittings (SKS phase) and surface deformation. As shown in New-Zealand[22] and California 2012[21], the comparison between surface deformation rate measured by continental (20–50 km spacing) GPS networks and shear-wave splittings measurements allows the comparison between current average lithosphere deformation and vertically cumulated strain as sampled by seismic waves (SKS and SKKS) through, asthenosphere and lithosphere. In western and central Alps, the presence of Lattice Preferred Orientations (LPO) anisotropy (delays $\delta t$ up to ~2 s) has been detected[63]. Amplitudes of travel time delays leave little doubt about their origin. Cumulating such large time delays over a wide region while keeping their orientation stable (only +/−17 degrees of variation within Switzerland) would require a strong organization of the deformation of crustal layers in all directions, which is not observed. We conclude then that the strain mapped using SKS splitting's originates mostly from mantle deformation and the contribution of crustal anisotropy must be limited. However, because of the inherent difficulty to date strain, it is not possible to determine whether the asthenosphere flow was still active and/or only recent using shear-wave splittings only[63]. Here we have an opportunity to solve this problem by comparing fast axis of shear wave splitting with seismicity and strain/stress fields measured today. South of the *HN*, the shear-wave splitting directions (38 +/−17 degree north) are consistent with both the orientations of *T*- axes (26 +/− 25 degree north) but less with directions of maximum extension (−5 +/− 49 degree north). This suggests that 1) a significant amount of finite strain is present through the Alpine arc (they are ~40 degrees off to the GPS extension rates). North of the *HN* in the Neuchatel area, the directions of fast axes of shear wave splitting's align with directions of maximum shortening (Fig. 8a and b); implying either decoupling at the base of the lithosphere (or within the crust) exists or that the surface deformation observed is of shallow origin.

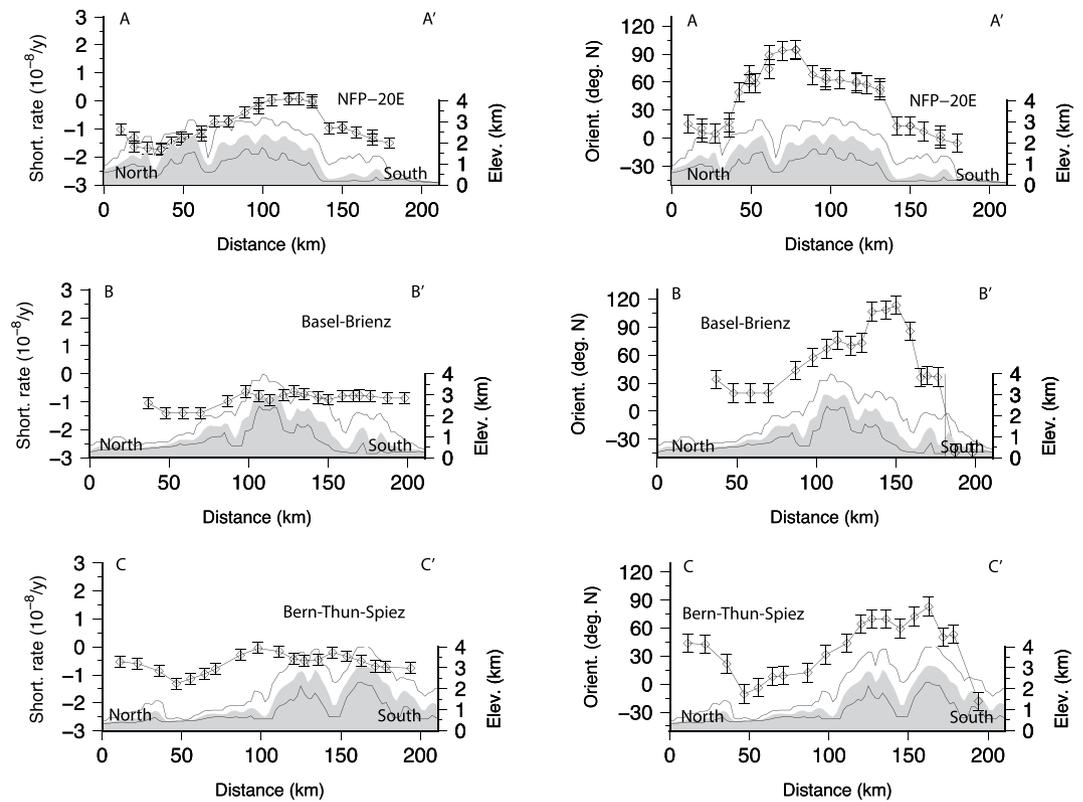

**Figure 6.** (**a**) Directions of maximum shortening as shown in Fig. 4b. (**b**) Maximum shortening rates along the profiles shown in Fig. 5. This figure has been created using Generic Mapping Tools 4/5[99] and Adobe Illustrator CS3.

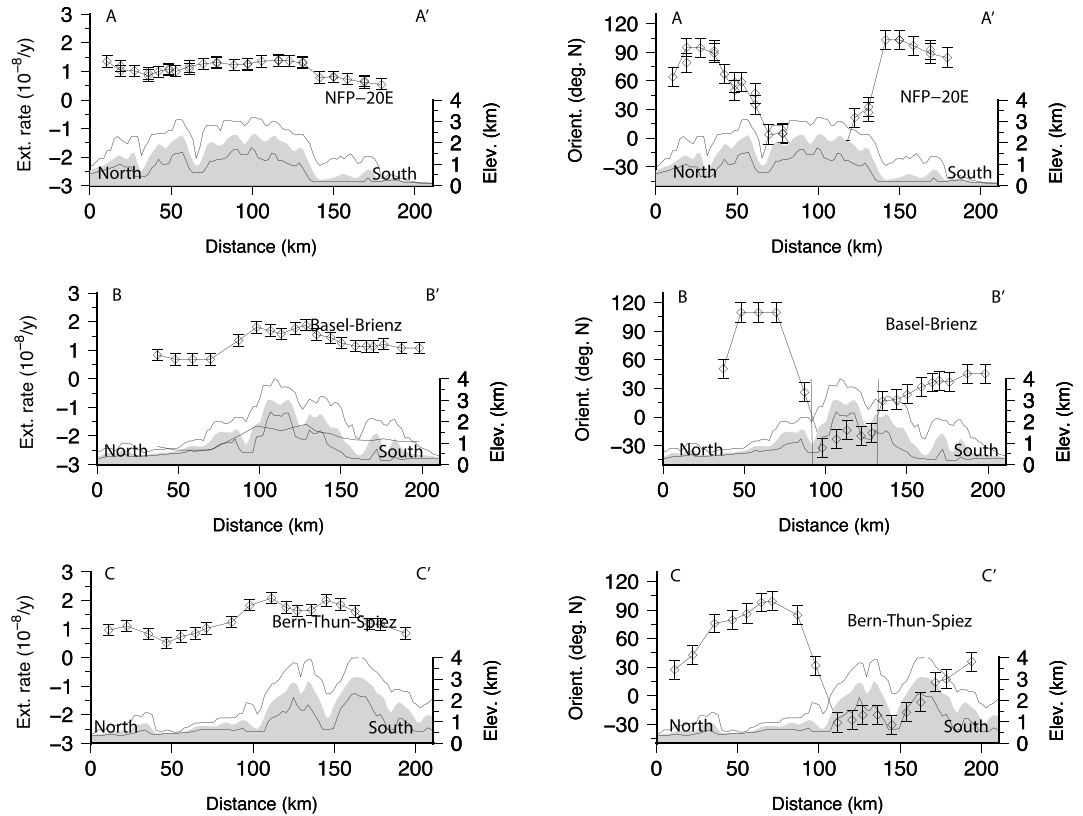

**Figure 7.** (**a**) Directions of maximum extension as shown in Fig. 4b; (**b**) Maximum extension rates along the profiles shown in Fig. 5.

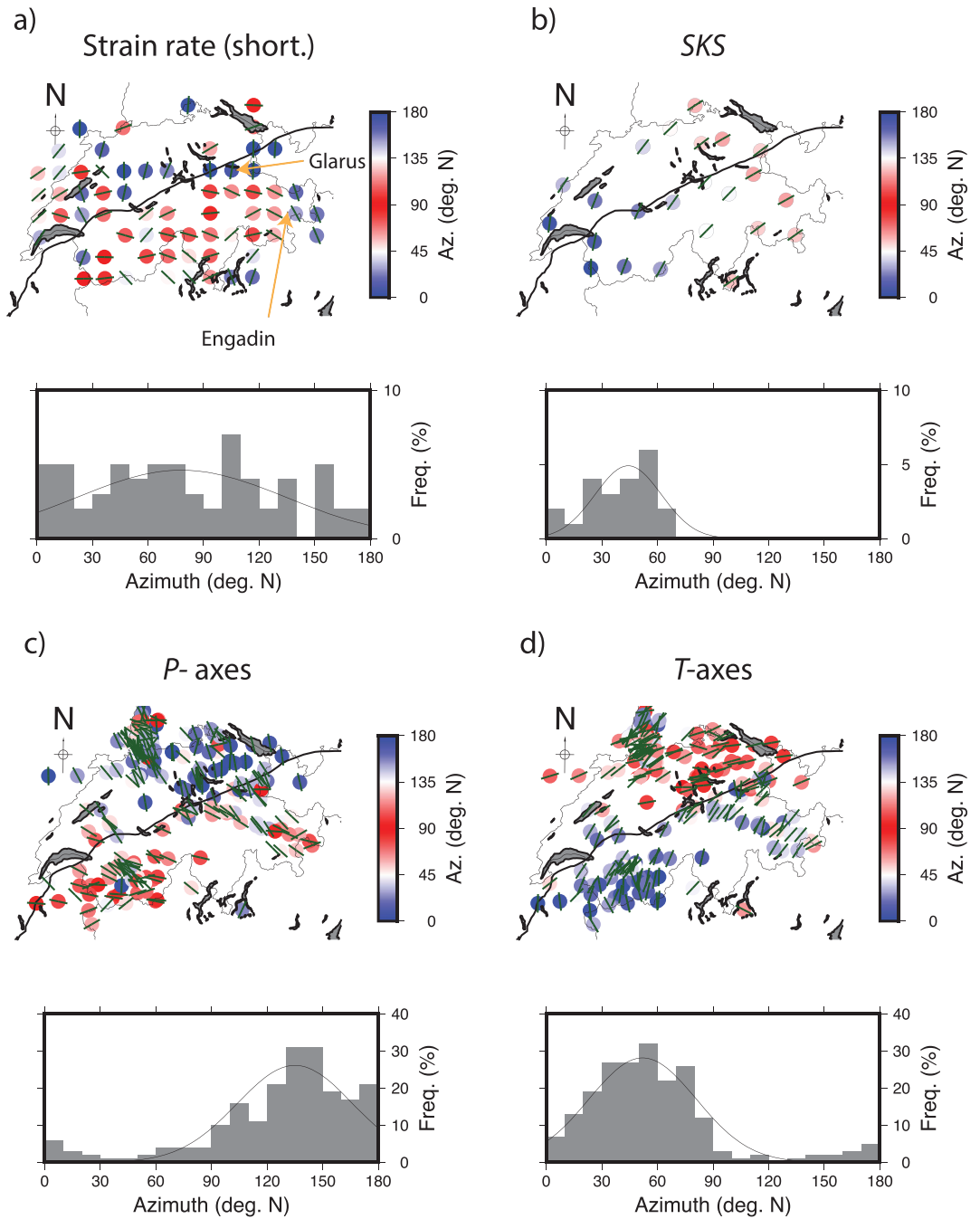

**Figure 8.** (**a**) Orientation of maximum shortening in Switzerland. South of the Alpine front shortening is oriented in the ~EW direction (N90) but cannot be determined in the Molasses, South of the Alpine front, in the Jura area, the picture remains more complex. The limits of the Alpine front is not sufficient to explain why in the east Engadin area the shortening orientations are 90 degrees (at ~N180) to the orientations observed anywhere else south of the Alpine front. (**b**) Fast axis of shear wave splitting[63]. GPS shortening directions rotate across the Alpine front while fast axis of shear wave splittings not. Unlike the other datasets, the orientation of the fast axes of shear wave splitting's do not rotate. In many places, the orientations of T- axes are sub-parallel to the fast axes of shear wave splitting's. (**c**) Orientation of P-Axes in Switzerland, north and south of the Alpine front. A 30 degrees (anti-clockwise) rotation is observed across the Alpine front line. North of the Alpine front, P- axes are ~perpendicular to the Alpine front, in agreement with the orientation of the $S_H$ in the Molasses[103]. (**d**) Orientation of T-Axes in Switzerland. In most Switzerland, T-axis are sub-parallel to the Alpine front. For each panel, we show the statistical distribution of each dataset plotted in the 0–180 degrees range. Mean orientations for directions of maximum shortening, P-, T- axes and fast axes of shear wave splittings are ~N80, ~N140, ~N50 and N37 respectively.

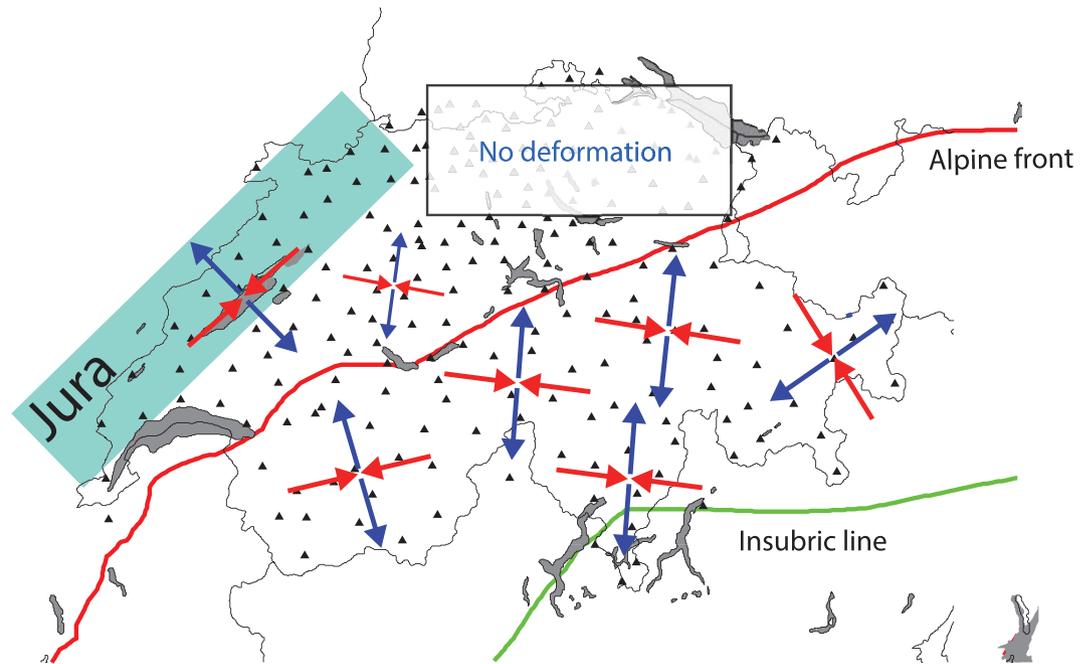

**Figure 9.** Summary sketch showing deformation in the study area.

## Conclusion

Switzerland represents a transition between stable continental regions with infrequent moderate to large magnitude events ($M_w < 6$) and more active areas where mainshocks ($M_w > 6$) occur every 50 years or less. We compare GPS surface strain rate field, shear wave splitting measurements, and principal components of crustal stress (*P*-/*T*- axes) of Switzerland in order to get a clearer picture of the processes active today. Each dataset gives access to various time- (from $10^2$ yr to $>10^5$ yr) and space-scales (from $10^4$ m to $10^5$ m) that we compare in the same framework.

At the country scale, seismicity and lithosphere strain rates are compatible through their respective orientations. We find the HN to be a major discontinuity that allows mapping both the changes of orientations of principal components of strain-rate and stress tensors and the changes in the seismicity depths (Fig. 9). The agreement of long- and short-space scale measurement seems to suggest that the central Alps are not close to experience a rapid extensional collapse such as described elsewhere[86,87]. The Jura, however, seems to be the place of a current geologic activity (i.e. extension combined with subsidence) that will require more investigations to be better understood.

Strain accumulated in the Valais and the Ticino are of similar amplitudes (~2 $10^{-8}$/yr), while the seismic moment released in the Ticino is ten times smaller than in the Valais and may therefore either 1) lead to an event of significant magnitude that is not present in today's historical catalogue for this area or 2) could be explained by the occurrence of dislocation creep. within the lithosphere.

Albeit we suggest that strain rates constrained from surface measurements fit well with the orientations of the stress components in the crust during the period covered by seismicity catalogues, we stay aware that the current lithosphere seismic activity as we measure it today may be well not representative of longer periods of time; the clustered seismicity observed in Switzerland could well suggest that even if slip rates are small today, they may vary over time periods as it has been observed in more active areas[16,88,89].

| Network solutions | Campaigns / Stations | Date |
|---|---|---|
| LV95 | 24 campaigns, 287 points | 1988–1995 |
| LV95 densification | 32 campaigns, 134 points | 1995–1998 |
| CHTRF98++ | 8 campaigns, 215 points | 1998–2002 |
| EUVN97 | 1 campaign, 217 points | 1997 |
| AGNES | 4.5 years permanent observed, 85 points | 1998–2003 |

**Table 2.** Description of the GPS campaigns[105] carried out by SWISSTOPO corresponding to vectors shown in Fig. 1b.

| Catalogue | Number of events | Average seismic Moment ($M_0$ in $10^{12}$/yr) | Period covered | Reference |
|---|---|---|---|---|
| First motions | 141 | 0.36 +/− 1.12 | 1968–1999 | [106] |
| First motions | 211 | 0.27 +/− 0.6 | 1968–2013 | [107] |
| Magnitude | 9559 | 0.52 +/− 0.5 | 1991–2008 | [37] |

**Table 3.** Seismicity catalogues used in this study. Seismic moment rates are shown in Fig. 9.

## Materials and Methods

**A-Datasets.** We use two datasets that have their own spatial and temporal resolutions.

- First, we use a catalogue of moment tensor solutions (first motions) for Switzerland ($M_w < 5$) to inform us on the orientation of the stress field through the analysis of *P*- and *T*- axes orientations. Principal axes of the moment tensor solutions give approximate orientation of the local stress field in the crust and span over the last ~40 years.
- Second, we use the strain rates computed from GPS velocity fields. Because of the GPS network density (<1 site per $10^3$ km$^2$) interpolation scheme used in this paper (see later for a full description), we are not able to resolve strain rates generated by small fault systems and any shallow sub-surface processes (<5–10 km depth). The GPS strain rate field has then very much to do with lithosphere deformation and less with crust structure. The agreement between stress and strain rate tensors principal components orientations would indicate that crustal and lithosphere deformation are consistent. Two things could mask this agreement: existing fault systems and post-seismic activity. Inherited fault systems can drive the orientation of rupture along fault, orientation that may not necessarily consistent with the stress inferred of borehole breakouts.

The GPS velocity field, results of campaigns carried out over the last 2 decades. We expect the surface velocities to be representative of the interseismic deformation field plus a component of postseismic deformation. Regarding postseismic activity, thanks to studies carried out on the postseismic deformation period that followed the 2004 $M_w$6 Parkfield[90] and the 2009 $M_w$6.3 L'Aquila[91] earthquakes, we know that the postseismic deformation period that follows a $M_w$6.0 earthquake does not last for more than 15 months. In consequence, as the last $M_w$ ~ 6.0 event in Switzerland occurred in Sierre in 1846, we assume the surface velocity field observed today is not disturbed by large scale postseismic deformation. We expect then that the velocity field observed today is representative of a period that could be longer than the time interval between the oldest and the most recent campaigns.

*Geodetic data.* We use the GPS surface velocity field (CHTRFv10) resulting of the processing of GPS campaigns (Table 2) and continuous GPS network data by SWISSTOPO teams[68,69,92]. Data collected in Switzerland have been processed using sites located in Europe (GRAZ, PFA2, PFAN, WETT and ZIM2) allowing a more stable estimation of troposphere state and of the orbits parameters. Formal uncertainties on velocities (<$10^{-5}$ m/yr) tells us that the velocity field's estimation is robust but do not reflect well the day to day repeatabilities (~3 mm) usually observed on time-series of permanent sites; even troposphere conditions are well constrained[93]. For the rest of the study, and in the light of campaign characteristics (first, last epoch), we assume that uncertainties on horizontal velocities components are of 0.1 mm/yr.

*Seismic data.* The Swiss Seismological Service (SED, http://www.seismo.ethz.ch/index_EN) reports local seismic activity (locations, magnitudes and when possible moment tensors) and routinely estimates moment tensors and earthquake focal mechanisms. This effort resulted in an accurate catalogue of *P*- and *T*- axes[81,94] made of 211 events located in Switzerland and surrounding areas for the period 1968–2013 (Fig. 2 and Table 3).

As these events represent only a fraction of the seismic moment released during the interseismic period, we supplement the catalogue of moment tensors with the earthquake catalogue ECOS-09 that enable us to map the spatial distribution of the seismic activity (Fig. 2c). The ECOS09 catalogue (http://hitseddb.ethz.ch:8080/ecos09/

introduction.html) is complete down to intensity V since 1878, IV since 1964 and Mw3.0 since 1976[37]. We list in Table 3 the characteristics of each earthquake catalogue mentioned above.

**B- Strain rate computation method.** We completed an inversion of GPS horizontal velocities to constrain the amplitudes and directions of the horizontal strain rate components using the software SSPX version 2D[95]. The strain rate field has been computed on a regular grid using a nearest-neighbour approach (node interval of 25 km and 6 neighbouring GPS sites included at each grid point in case they are within 50 km of the node). We assume that strain rates are not significant when rates computed from uncertainties on velocities are larger than rates computed using their amplitudes. In other words, at a specific location, if the strain rate computed from uncertainties on velocities is larger than the ones computed using velocity amplitudes, the strain rate at this location is set to zero. This approach, however, does not prevent to be more selective: we consider that any strain rate smaller than $2.10^{-9}$ strain/yr (or a relative motion of 2 mm/yr over 1000 km) should not be trusted.

### Acknowledgements

We thank Pr. E. Kissling, Pr. S. Schmid, Dr. A. Villiger, Dr. V. Picotti and the writing club group of SEG for their contributions in improving the manuscript. We also thank Pr. Dr. Niu and Pr. Dr. Schlunegger plus two anonymous reviewers. This work could not have been possible without the support of the ARC1/ARC2 team of the University of Leeds, UK.

### Author Contributions

N.H. initial idea; N.H. and J.W. completed the figures design; N.H., J.W., D.G., and M.R. wrote the text.

### Additional Information

**Supplementary information** accompanies this paper at https://doi.org/10.1038/s41598-018-20253-z.

**Competing Interests:** The authors declare that they have no competing interests.

**Publisher's note:** Springer Nature remains neutral with regard to jurisdictional claims in published maps and institutional affiliations.